\documentclass[a4paper]{jpconf}
\usepackage{graphicx}
\usepackage{iopams}

\begin{document}

\title{Viscosities of the quasigluon plasma}

\author{M Bluhm$^{1}$, B K\"ampfer$^{2,3}$ and K Redlich$^4$}

\address{$^1$ SUBATECH, UMR 6457, Universit\'{e} de Nantes, 
Ecole des Mines de Nantes, IN2P3/CRNS. 4~rue Alfred Kastler, 
44307 Nantes cedex 3, France}
\address{$^2$ Institut f\"ur Strahlenphysik, 
Forschungszentrum Dresden-Rossendorf, PF 510119, 01314 Dresden, 
Germany}
\address{$^3$ Institut f\"ur Theoretische Physik, TU Dresden, 
01062 Dresden, Germany}
\address{$^4$ Institute of Theoretical Physics, University of 
Wroclaw, 50204 Wroclaw, Poland}

\ead{bluhm@subatech.in2p3.fr}

\begin{abstract}
We investigate bulk and shear viscosities of the gluon plasma 
within relaxation time approximation to an effective 
Boltzmann-Vlasov type 
kinetic theory by viewing the plasma as describable in terms 
of quasigluon excitations with temperature dependent 
self-energies. The found temperature dependence of the transport 
coefficients agrees fairly well with available lattice QCD results. 
The impact of some details in the quasigluon dispersion relation 
on the specific shear viscosity is discussed. 
\end{abstract}

\section{Introduction \label{sec:1}}

Bulk $\zeta$ and shear $\eta$ viscosities are important properties of 
strongly interacting matter. Their firm knowledge is a necessary 
prerequisite for dynamical descriptions of relativistic heavy-ion 
collisions, but also in cosmology and astrophysics. The success of 
ideal relativistic hydrodynamic simulations, 
cf.~e.~g.~\cite{Kolb04}, in quantitatively describing the 
collective flow behaviour of the matter produced at the Relativistic 
Heavy Ion Collider led to the conclusion that at most small dissipative 
effects are present in the created medium~\cite{Heinz02,Lacey07,Drescher07}. 
Later investigations by means of dissipative relativistic hydrodynamics, 
cf.~e.~g.~\cite{Romatschke07,Dusling08,Song08,Luzum08}, confirmed 
this observation. 

However, the quantum mechanical uncertainty principle enforces a 
fundamental lower bound on the shear viscosity~\cite{Danielewicz85} for 
any physical system. Moreover, unitarity arguments led to a lower limit 
conjecture for the specific shear viscosity~\cite{Policastro01,Kovtun03} 
$(\eta/s)_{KSS}\ge \hbar/(4\pi)$ (Kovtun-Son-Starinets bound, 
$s$ is the entropy density). The bulk viscosity, 
in contrast, is exactly zero in conformal theories. In QCD, however, 
it is only (approximately) zero at large temperatures $T$, while for $T$ 
close to the deconfinement transition temperature $T_c$ the specific bulk 
viscosity $\zeta/s$ is expected to become large~\cite{Kharzeev08,Karsch08,Romatschke09}. 

The viscosity coefficients were recently evaluated by means of non-perturbative 
first-principle numerical lattice QCD 
calculations~\cite{Nakamura05,Sakai07,Meyer07,Meyer08}. Other analytic approaches 
are based on the Kubo-formalism~\cite{Jeon95} or on linearized Boltzmann 
kinetic theory~\cite{Jeon95,Jeon96,Gagnon07-1,Gagnon07}. In fact, it was shown 
that in scalar theory~\cite{Jeon95,Jeon96} and in hot 
QED~\cite{Gagnon07-1,Gagnon07} both approaches are equivalent. For 
small QCD running coupling, perturbative QCD (pQCD) kinetic theory results at 
leading order for shear and bulk viscosities were reported in~\cite{Arnold00,Arnold03} 
and~\cite{Arnold06}, respectively. 

\section{Bulk and shear viscosity coefficients \label{sec:2}}

We calculate bulk and shear viscosities of the gluon plasma by assuming 
that it can be described by quasigluon excitations with $T$-dependent self-energy 
$\Pi(T)$, cf.~\cite{Bluhm09,Bluhm10}. In local thermal equilibrium, this picture 
is comprised within a quasiparticle model (QPM), where the gluon 
dispersion relation reads $E^0=\sqrt{\vec{p}^{\,2}+\Pi(T)}$. 
$\Pi(T)=T^2 G^2(T)/2$ contains the effective coupling 
$G^2(T)=16\pi^2/\left(11\ln\left[\lambda(T-T_s)/T_c\right]^2\right)$ 
with parameters $\lambda$ and $T_s$. The QPM works 
impressively well for describing equilibrium lattice QCD thermodynamics, 
cf.~\cite{Bluhm05,Bluhm07-2,Bluhm08-1}. 

By means of an effective kinetic theory~\cite{Jeon95,Jeon96}, the model can be 
extended to systems facing small deviations from local thermal equilibrium. 
In fact, as $T$ is space-time dependent in general, so is the dispersion 
relation $E(x)$. The corresponding energy-momentum tensor reads 
$T^{\mu\nu}(x) = \int \frac{d^3 \vec{p}}{(2\pi)^3E(x)} p^\mu(x)p^\nu(x) b(x,p) 
+ g^{\mu\nu} B(\Pi(x))$, where $b(x,p)$ represents the quasigluon distribution 
function entering $E(x)$ and $p^\alpha=(E,\vec{p}\,)$. This 
expression for $T^{\mu\nu}$ is interwoven with a Boltzmann-Vlasov type kinetic 
equation~\cite{Jeon95}. The mean field term $B(\Pi)$ is necessary for 
satisfying locally energy and momentum conservation, 
$\partial_\mu T^{\mu\nu} = 0$, established when~\cite{Bluhm09,Bluhm10} 
$\partial B/\partial\Pi=-\frac12\int \frac{d^3 \vec{p}}{(2\pi)^3E(x)} b(x,p)$. 
This also guarantees thermodynamic self-consistency of the approach in thermal 
equilibrium and $E=\delta T^{00}/\delta b$ as a basic principle of statistical 
mechanics. 

Expanding $b(x,p)=b^0(x,p)+\delta b(x,p)$ for small disturbances $\delta b$ 
from equilibrium $b^0$, $T^{\mu\nu}$ is decomposed in an equilibrium part and 
first-order corrections thereof. Using the relaxation time approximation for 
determining $\delta b$ in the latter, one reads off~\cite{Bluhm10} 
for $\eta$ and $\zeta$ in the local fluid rest frame by comparison with the 
phenomenological definitions~\cite{deGroot} 
\begin{eqnarray}
 \label{equ:eta2} 
 \eta & = & \frac{1}{15T} \int \frac{d^3 p}{(2\pi)^3} 
 n(T)[1+d^{-1}n(T)] \frac{\tau}{(E^0)^2}\vec{p}^{\,4} \,, \\
 \label{equ:zeta3}
 \zeta & = & \frac{1}{T} \int \frac{d^3 p}{(2\pi)^3} 
 n(T)[1+d^{-1}n(T)] \frac{\tau}{(E^0)^2} 
 \left\{ \left[(E^0)^2-T^2 \frac{\partial\Pi}{\partial T^2}
 \right]\frac{\partial P}{\partial\epsilon} 
 - \frac13 \vec{p}^{\,2} \right\}^2 \,,
\end{eqnarray}
where $n(T)=d(e^{E^0/T}-1)^{-1}$ with degeneracy factor $d=16$, 
$\partial P/\partial\epsilon$ is the squared speed of sound and $\tau$ the relaxation 
time. These results generalize previous work~\cite{Hosoya85,Gavin85}, in which 
excitations with constant mass $M$ were considered. While the expression for 
$\eta$ looks formally the same, cf.~also~\cite{Sasaki09-1}, $\zeta$ 
in Eq.~(\ref{equ:zeta3}) contains a combination of $E^0$ and self-energy 
derivative in contrast to~\cite{Hosoya85,Gavin85,Sasaki09-1}. It is this 
particular combination which guarantees $\zeta\rightarrow 0$ in the conformal 
limit. 

\section{Numerical results \label{sec:3}}

\begin{figure}[ht]
\includegraphics[scale=0.31]{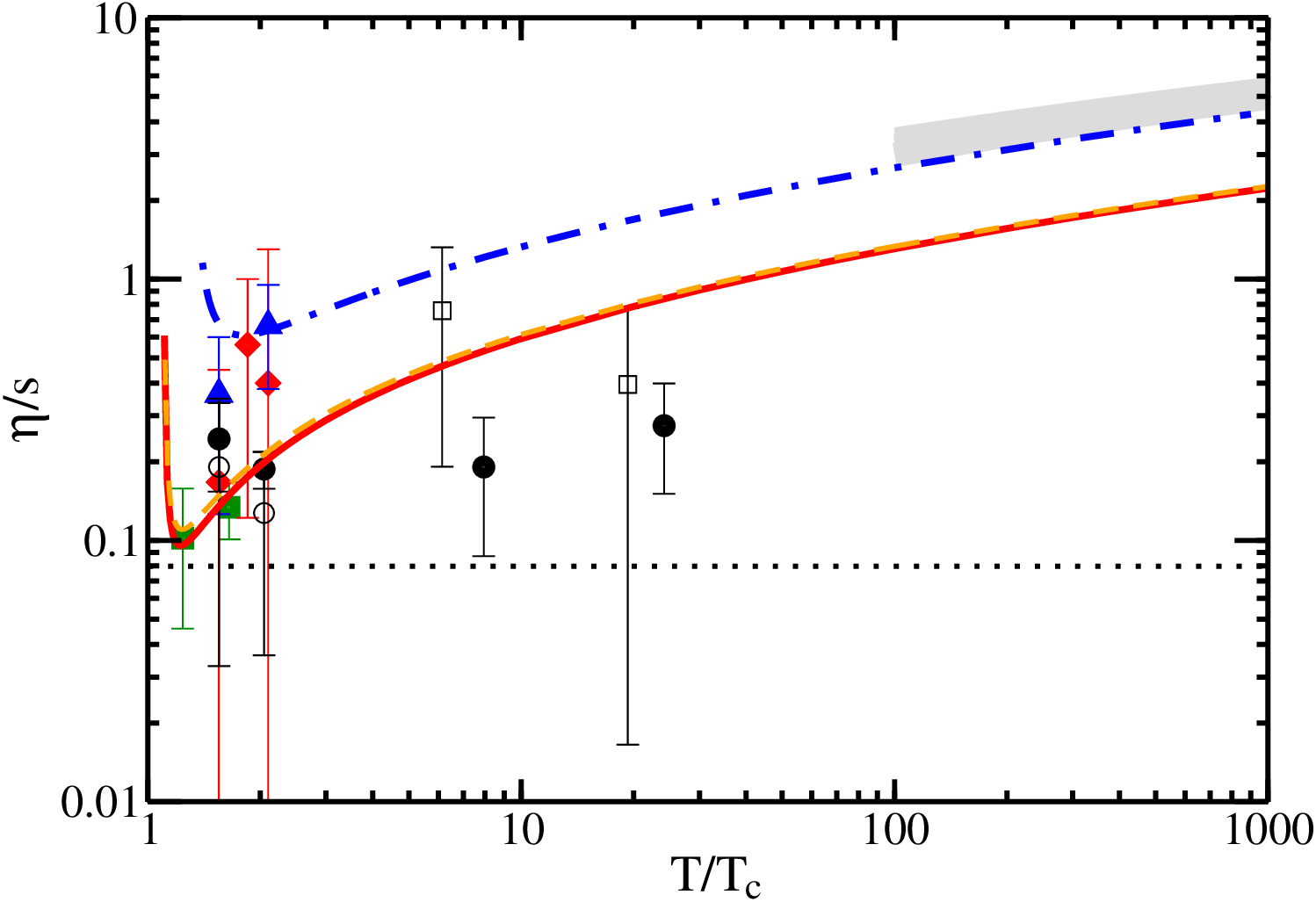}
\hspace{3mm}
\includegraphics[scale=0.31]{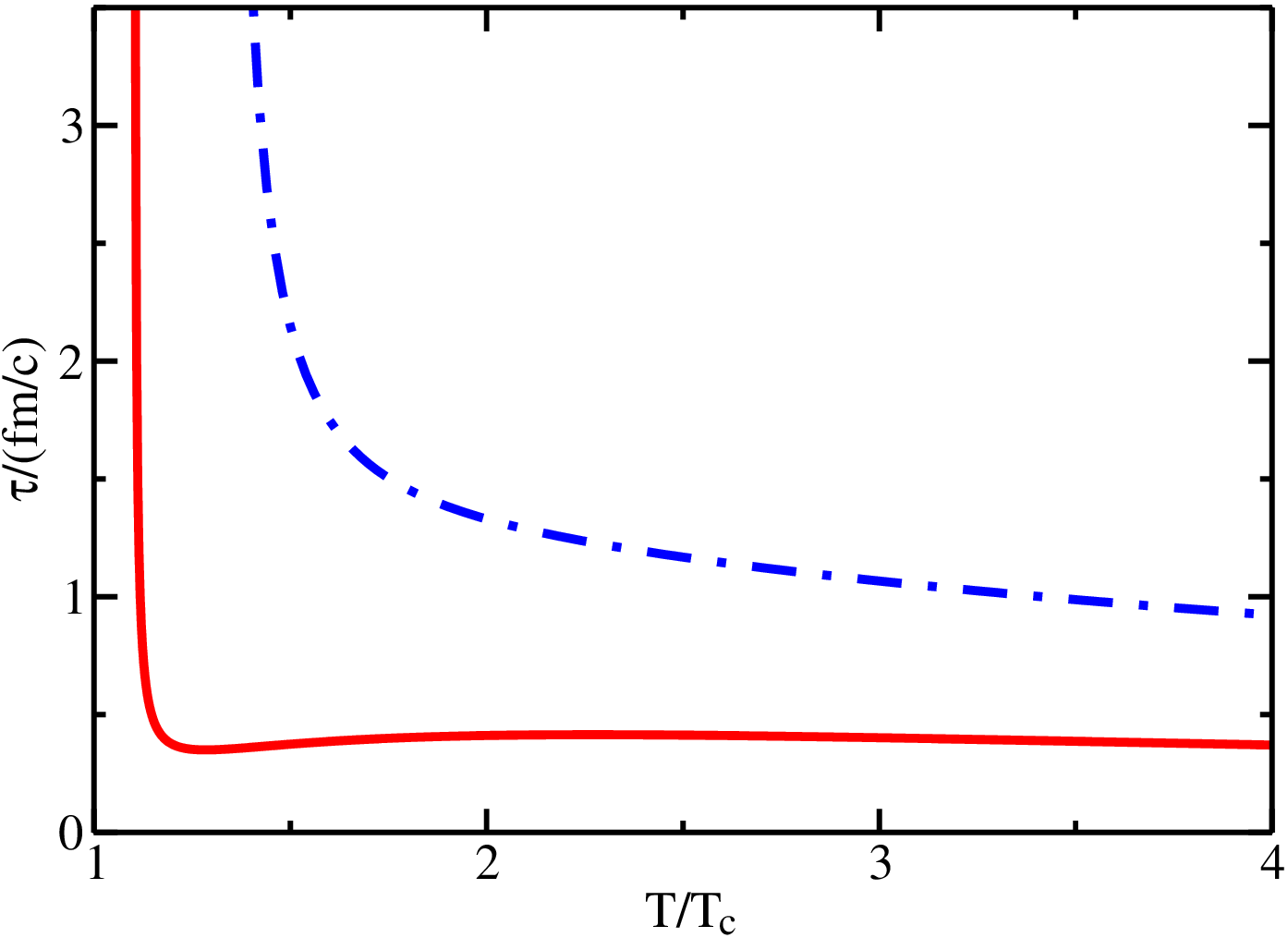}
\caption{\label{fig:etas+relax} Left: Specific shear viscosity as a 
function of $T/T_c$. Dash-dotted and solid curves exhibit QPM results 
for Fit~1 and Fit~2, respectively, compared with lattice QCD data 
from~\cite{Meyer07} (full squares),~\cite{Nakamura05} (diamonds and 
triangles) and~\cite{Sakai07} (open squares, open and full circles). 
The dotted line exhibits the KSS bound. The grey band at 
large $T$ depicts the pQCD result 
$\eta_{NLL}/\tilde{s}$, where $\eta_{NLL}$ is from~\cite{Arnold03} and 
$\tilde{s}$ is from~\cite{Blaizot01}. 
The dashed curve (almost on top of the solid curve) shows $\eta/s$ for 
Fit~2 for constant mass $M=200$ MeV instead of $\Pi(T)$ in $E^0$. 
Right: Corresponding relaxation times $\tau$. The behaviour is crucially 
determined by the parameter $a_2$.}
\vspace{-4mm}
\end{figure}

\begin{figure}[ht]
\vspace{0.45cm}
\includegraphics[scale=0.31]{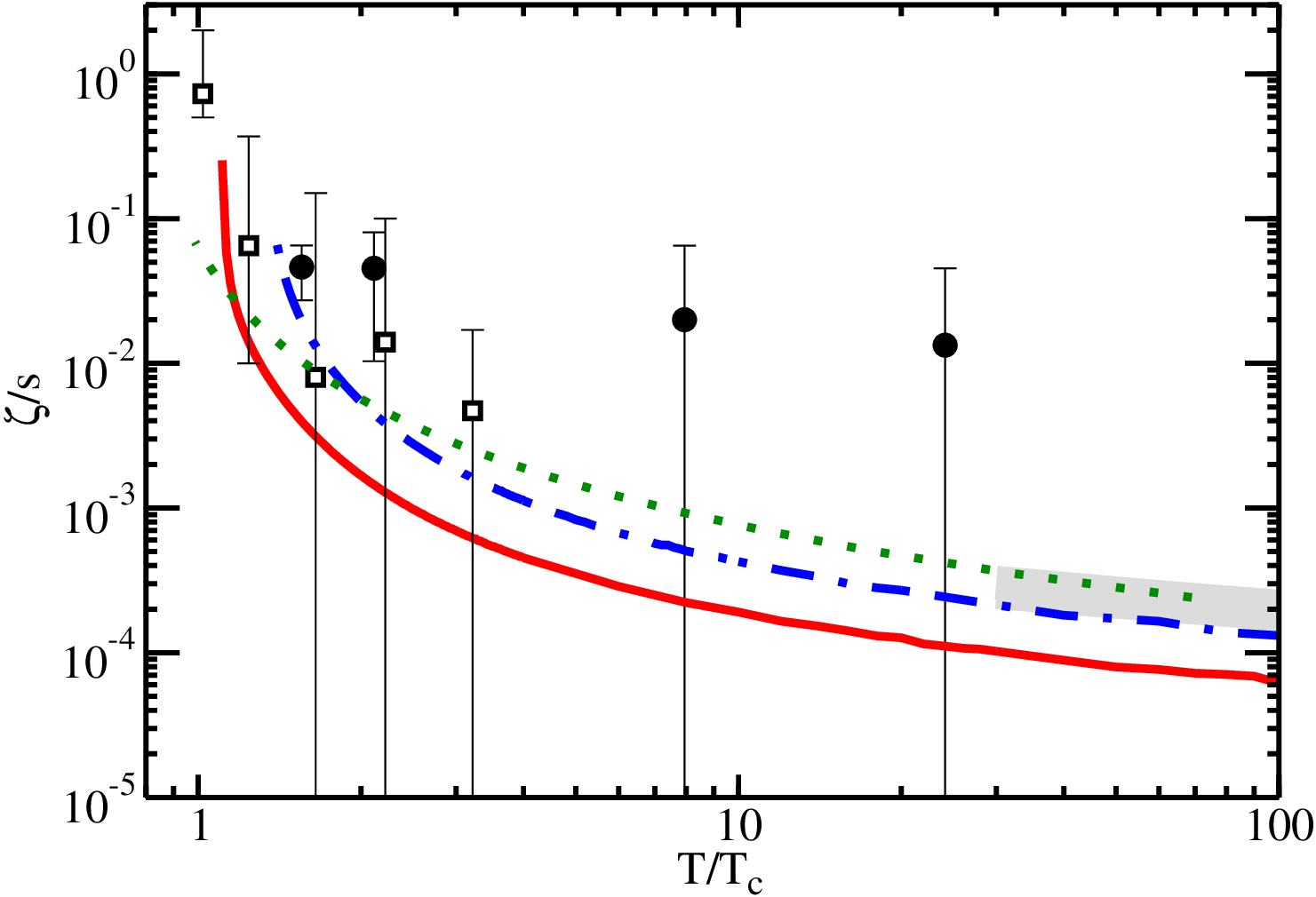}\hspace{2pc}%
\begin{minipage}[h]{15pc} \vspace{-5.8cm}
\caption{\label{fig:zetas} Specific bulk viscosity as a function of 
$T/T_c$. Dash-dotted and solid curves exhibit QPM results 
for Fit~1 and Fit~2, respectively, 
compared with available lattice QCD data from~\cite{Meyer08} 
(squares) and~\cite{Sakai07} (circles). 
The dotted curve shows for comparison results from 
holographic QCD~\cite{Guersoy09}. The grey band at large $T$ depicts 
the pQCD result $\zeta_{LO}/\tilde{s}$, where $\zeta_{LO}$ 
is from~\cite{Arnold06} and $\tilde{s}$ is from~\cite{Blaizot01}.}
\end{minipage}
\vspace{-4mm}
\end{figure}

For numerically evaluating Eqs.~(\ref{equ:eta2}) and~(\ref{equ:zeta3}), we 
employ as ansatz for the relaxation time $\tau = [a_1 T G^4(T) \ln (a_2/G^2(T))]^{-1}$, 
which is inspired by considerations in~\cite{Heiselberg94}. It turns out that with 
this ansatz for $\tau$, both $\eta$ and $\zeta$ resemble at large $T$ the behaviour 
with temperature and coupling known from pQCD~\cite{Arnold00,Arnold03,Arnold06}. 
The parameters in $G^2(T)$ are adjusted such that thermal equilibrium lattice QCD 
results for $s/T^3$ from~\cite{Boyd96,Okamoto99} are described by the QPM. Local 
thermal equilibrium quantities, such as $s$, are obtained by inserting $n(T)$ for 
$b^0(x,p)$ into $T^{\mu\nu}(x)$, cf.~\cite{Bluhm10}. 
Furthermore, we fit the parameters in $\tau$ as $a_1=2.587\times 10^{-2}$ and 
$a_2=(\mu_*/T)^2$ (Fit~1) or as $a_1=3.85\times 10^{-2}$ and 
$a_2=2(\mu_*/T)^2$ (Fit~2), where 
$\mu_*/T=2.765$ as in pQCD~\cite{Arnold03}. QPM results 
for $\eta/s$ and the corresponding relaxation times are exhibited 
in Fig.~\ref{fig:etas+relax}, while results for $\zeta/s$ are depicted in 
Fig.~\ref{fig:zetas}. 

Good agreement with available lattice QCD results is found. In particular, 
$\eta/s$ exhibits a minimum in the vicinity of $T_c$, which is common for a variety 
of liquids and gases~\cite{Kovtun03,Csernai06}. We find $(\eta/s)_{min}^*=0.096$ 
at $T=1.22\,T_c$ for Fit~2. The exact location of the minimum is driven by 
$a_2$ in $\tau$. With increasing $T$, $\eta/s$ increases rather mildly, 
while for $T\rightarrow T_c^+$ it dramatically rises in line with the relaxation 
time and the phenomenon of critical slowing down. 

For comparison, $\eta/s$ for Fit~2 based on Eq.~(\ref{equ:eta2}) but for 
particles with constant $T$-independent mass $M=200$~MeV is also 
exhibited in Fig.~\ref{fig:etas+relax} (dashed curve). Only the region around 
the minimum in $\eta/s$ is affected (the minimal value is about $14\,\%$ 
larger than $(\eta/s)_{min}^*$), while at large $T$ details in $E^0$ become 
irrelevant. Increasing $M$ generally leads to a decrease in $\eta/s$. For 
$M=1$ GeV a minimal value $9\,\%$ smaller than $(\eta/s)_{min}^*$ is found. 

\section{Conclusion \label{sec:4}}

We derived expressions for bulk and shear viscosities within relaxation time 
approximation to an effective kinetic theory for the gluon plasma assumed to be 
composed of quasigluon excitations with $T$-dependent dispersion 
relation $E^0$. Medium effects show up in $\eta$ only implicitly via $E^0$, 
while they are prominently evident in $\zeta$. 
Specifying the relaxation time and adjusting model parameters to thermal 
equilibrium lattice QCD results, we find good agreement with available 
lattice QCD data for $\eta/s$ and $\zeta/s$ for $T$ close to $T_c$. The influence of 
moderate constant masses $M$ on $\eta/s$ is observed to be small and focussed on 
the region around its minimum. At large $T$ our results resemble the parametric 
dependence on $T$ and coupling known from pQCD. 

\ack
We acknowledge insightful discussions with S.~Jeon, 
U. G\"ursoy, E. Kiritsis, H.~B.~Meyer, S.~Peigne, 
A.~Peshier, C.~Sasaki and U.~Wiedemann. 
M.B. and K.R.~acknowledge the hospitality of 
the CERN TH department. This work is supported by 
06 DR 9059, GSI-FE, the European Network I3-HP2 Toric 
and the Polish Ministry of Science.

\end{document}